\documentclass[epj]{svjour}
\usepackage{graphics}
\begin{document}
\title{Dynamics of a superconducting qubit coupled to the quantized cavity field: a unitary transformation approach}
\author{Dagoberto S. Freitas\inst{1,2}\thanks{Permanent address: Departamento de F\'\i sica, Universidade Estadual de
Feira de Santana, 44036-900, Feira de Santana, BA, Brazil} 
\and A. Vidiella-Barranco\inst{3} \and J.A Roversi\inst{3}
%
}                     
\offprints{A. Vidiella-Barranco}          
%
\institute{
Departamento de F\'\i sica,
Instituto de Ci\^encias Exatas, Universidade Federal de Minas
Gerais, 30123-970, Belo Horizonte, MG, Brazil\and
Departamento de F\'\i sica, Universidade Estadual de
Feira de Santana, 44036-900, Feira de Santana, BA, Brazil\and
Instituto de F\'\i sica ``Gleb Wataghin'', Universidade
Estadual de Campinas, 13083-859 Campinas,  SP, Brazil }
\date{Received: date / Revised version: date}
%
\abstract{
We present a novel approach for studying the dynamics of a
superconducting qubit in a cavity. We succeed in linearizing the
Hamiltonian through the application of an appropriate unitary
transformation followed by a rotating wave approximation (RWA). For
certain values of the parameters involved, we show that it is
possible to obtain a a Jaynes-Cummings type Hamiltonian. 
As an example, we show the existence of super-revivals for the 
qubit inversion.
} 
\titlerunning{New approach to the dynamics of a superconducting qubit...}
\maketitle
\section{Introduction} Superconducting qubits,
which consist of a small superconducting electrode connected to a
reservoir via a Josephson junction \cite{yu01}, are considered to be
promising qubits for quantum information processing. Because of the
charging effect in the small electrode, the two charge-number
states, in which the number of Cooper pairs in the``box" electrode
differs by one, constitute an effective two-level system.
These ``artificial atoms," with well-defined discrete energy levels,
provide a platform to test fundamental quantum effects, e.g.,
related to cavity quantum electrodynamics (cavity QED) as well as
for quantum information schemes. A particularly interesting proposal 
is a viable architecture for quantum computation based on circuit cavity 
QED, as presented in \cite{blais04}.
A few quantum features of that system have been already experimentally 
demonstrated some years ago; for instance, the generation of
superposition of the two charge states \cite{bouchiat98,nakamura97}, 
and coherent oscillations between two degenerate
states \cite{nakamura99}. More recently, an important step for 
reaching the quantum regime in such systems has been achieved: the strong 
coupling of the quantized radiation field to a superconducting qubit, as 
experimentally demonstrated by A. Wallraff et al. \cite{wallraff04}.
Accordingly, a superconducting qubit coupled to the quantized field
may be used to engineer quantum states; for instance, as a deterministic
single-photon source as well as to generate arbitrary superpositions of
Fock states of the cavity field, as proposed in \cite{yu04}. 
As a matter of fact, the generation of Fock 
(and coherent) states has been already performed in such
systems \cite{houck07,hofheinz08}. 
We may also cite as important developments involving superconducting systems 
appplied to quantum information schemes, the demonstrations of quantum buses
using superconducting qubits and photons \cite{sillanpaa07,majer07},
the implementation of two-qubit algorithms \cite{{dicarlo09}}, 
the encoding of quantum information in Schr\"odinger type cat states 
\cite{vlastakis13}, as well as the deterministic entanglement of
superconducting qubits \cite{riste13}. Moreover, it has also become possible 
to engineer an artificial nonlinear Kerr-type medium, allowing the 
observation of quantum state collapses and revivals \cite{kirchmair13}.  

In this paper we investigate a model of circuit QED following 
reference \cite{yu04}: 
a single mode cavity field in the microwave regime, with photon
transitions between the ground and first excited states of a
two-level system formed by a superconducting quantum
interference device (SQUID). This artificial two-level``atom" can be
easily controlled by an applied gate voltage $V_{g}$ and the flux
$\Phi_{c}$ generated by the classical magnetic field through the
SQUID.
Here we adopt a new approach to this problem: we depart from the
full radiation field and supercondutor interaction Hamiltonian, and
apply a unitary transformation similar to the one introduced in
\cite{moya}. After transforming the original Hamiltonian, we show
that it is possible to obtain a simpler Hamiltonian by performing
a rotating wave approximation (RWA); we are able to obtain a
Jaynes-Cummings like Hamiltonian which makes possible to solve the problem 
analytically without making any further approximations. In other words, 
we were able to linearize the superconductor/quantized field Hamiltonian without
doing the usual power series expansions of the Hamiltonian itself,
as usually found in the literature
\cite{yu04}.

\section{The model}
We consider a system constituted by a SQUID type superconducting box
with $n_{c}$ excess Cooper-pair charges connected to a
superconducting loop via two identical Josephson junctions having
capacitors $C_{J}$ and coupling energies $E_{J}$. An external
control voltage $V_{g}$ couples to the box via a capacitor $C_{g}$.
We also assume that the system operates in a regime, consistent with
most experiments involving charge qubits, in which only Cooper pairs
coherently tunnel in the junctions. Therefore the
system Hamiltonian may be written as \cite{yu01}

\begin{equation}
H_{qb} = 4E_{ch}(n_{c}-n_{g})^{2}-2E_{j}\cos(\frac{\pi
\Phi_{X}}{\Phi_{0}})\cos(\Theta),\label{Hqb}
\end{equation}
where $E_{ch} = e^{2}/2(C_{g} +2C_{J})$ is the single-electron
charging energy, $n_{g} = C_{g}V_{g}/2e$ is the dimensionless gate
charge (controlled by $V_{g}$), $\Phi_{X}$ is the total flux through
the SQUID loop and $\Phi_{0}$ the flux quantum. By adjusting the
flux through the superconducting loop, one may control the Josephson
coupling energy as well as switch on and off the qubit-field
interaction. The phase $\Theta = (\phi_{1}+\phi_{2})/2$ is the
quantum-mechanical conjugate of the number operator $n_{c}$ of the
Cooper pairs in the box, where $\phi_{i}$ (i = 1, 2) is the phase
difference for each junction. The superconducting box is assumed to
be working in the charging regime and the superconducting energy gap
$\Delta$ is considered to be the largest energy involved. Moreover,
the temperature $T$ is low enough so that the condition  $\Delta
\gg k_{B}T$ holds. The superconducting box then
becomes an effective two-level system with states $|g\rangle$ (for $n_{c} = 0$)
and $|e\rangle$ (for $n_{c} = 1$) given that the gate voltage is
near a degeneracy point ($n_{g} = 1/2$) \cite{yu01} and the
quasi-particle excitation is completely suppressed \cite{averin}.

If that circuit is placed within a single-mode microwave
superconducting cavity, the qubit can be coupled to both a classical
magnetic field (generates a flux $\Phi_{c}$) and the quantized
cavity field (generates a flux $\Phi_{q} =\eta a +\eta^{*}
a^{\dag}$, with $a$ and $a^{\dag}$ the annihilation and creation
operators), being the total flux through the SQUID $\Phi_{X} =
\Phi_{c} + \Phi_{q}$ \cite{you03}. The parameter
$\eta$ is related to the mode function of the cavity field. The
system Hamiltonian will then read

\begin{equation}
H = \hbar\omega a^{\dag}a + E_{z}\sigma_{z} - E_{J}(\sigma_{+} +
\sigma_{-})\cos\big(\gamma I + \beta a +\beta^{*}
a^{\dag}\big),\label{H}
\end{equation}
where we have defined the parameters $\gamma = \pi\Phi_{c}/\Phi_{0}$
and $\beta = \pi\eta/\Phi_{0}$. The first term corresponds to the
free cavity field with frequency $\omega = 4E_{ch}/\hbar$ and the
second one to the qubit having energy $E_{z} = -2E_{ch}(1 - 2n_{g})$
with $\sigma_z = |e\rangle\langle e| - |g\rangle\langle g|$. The
third term is the (nonlinear) photon-qubit interaction term which
may be controlled by the classical flux $\Phi_{c}$. In general the
Hamiltonian in equation (\ref{H}) is linearized under some kind of
assumption. In \cite{yu04}, for instance, the authors decomposed the
cosine in Eq.(\ref{H}) and expanded the terms $\sin[\pi(\eta\,
a+H.c.)/\Phi_{0}]$ and $\cos[\pi(\eta\, a+H.c.)/\Phi_{0}]$ as power
series in $a \,(a^{\dagger})$. In this way, if the limit $|\beta|\ll
1$ is taken, only single-photon transition terms in the expansion
are kept, and  a Jaynes-Cummings type Hamiltonian (JCM) is then
obtained. 
Here, in contrast to that, we adopt a similar technique to
the one presented in reference \cite{moya}; we obtain a linear, JCM-type
Hamiltonian by first applying a transformation to the full Hamiltonian
in equation (\ref{H}) and making approximations afterwards. 
The suitable unitary transformation is given by
\begin{eqnarray}
T &=& \frac{1}{\sqrt{2}} \Big\{-\frac{1}{2} \Big[
 D^{\dag}\Big(\alpha,\gamma \Big) -  D\Big(\alpha,\gamma \Big)
\Big]I \nonumber\\
&-&\frac{1}{2} \Big[
 D^{\dag}\Big(\alpha,\gamma\Big) + D\Big(\alpha,\gamma
\Big) \Big] \sigma_z\nonumber\\
&+& D\Big(\alpha,\gamma \Big)\sigma_{+} + D^{\dag}\Big(\alpha,\gamma
\Big) \sigma_{-} \Big\}, \label{T}
\end{eqnarray}
with
$D(\alpha,\gamma)=D(\alpha)e^{i\frac{\gamma}{2}}$, where $D(\alpha)
=\exp[(\alpha a^{\dag} -\alpha^{*}a)]$ is Glauber's displacement
operator, with $\alpha = i\beta^{*}/2$. We obtain the following
transformed Hamiltonian
\begin{eqnarray}
H_{T}  & \equiv &  THT^{\dagger}\nonumber\\
& = & \hbar\omega
a^{\dag}a+\frac{E_{J}}{2}\sigma_{z}+i\frac{\hbar}{2}\big[\omega\big(\beta
a- \beta^{*}a^{\dag}\big) +
2i\frac{E_{z}}{\hbar}\big]\big(\sigma_{+}+\sigma_{-}\big)\nonumber\\
& + & \frac{E_{J}}{2}\cos\big[2\big(\beta a+
\beta^{*}a^{\dag}\big)+2\gamma
\big]\sigma_{z}\nonumber\\
&-& i\frac{E_{J}}{2}\sin\big[2\big(\beta a+
\beta^{*}a^{\dag}\big)+2\gamma
\big]\big(\sigma_{+}-\sigma_{-}\big).\label{HT0}
\end{eqnarray}
It is worth mentioning that the same setup and transformation
given by (\ref{T}) may also be employed (see reference \cite{dago12})
in a scheme for preparation of superpositions of coherent states of a
single-mode cavity field (Schr\"{o}dinger cats), extending the approach
of Ref.\cite{yupra05}.

Now we rewrite $H_{T}$ in an interaction
representation in the transformed space, or
$H_{TI}=U_{0T}^{\dagger}V_{T}U_{0T}$, where
\begin{eqnarray}
V_{T} &=& i\frac{\hbar}{2}\big[\omega\big(\beta a -
\beta^{*}a^{\dag}\big) +
2i\frac{E_{z}}{\hbar}\big]\big(\sigma_{+}+\sigma_{-}\big)\nonumber\\
&+& \frac{E_{J}}{2}\cos\big[2\big(\beta a +
\beta^{*}a^{\dag}\big)+2\gamma
\big]\sigma_{z}\nonumber\\
&-& i\frac{E_{J}}{2}\sin\big[2\big(\beta a +
\beta^{*}a^{\dag}\big)+2\gamma
\big]\big(\sigma_{+}-\sigma_{-}\big)\nonumber\\
\end{eqnarray}
and $U_{0T} = e^{-i(\omega
a^{\dag}a+\frac{E_{J}}{2\hbar}\sigma_{z})t}$. We then obtain the
transformed Hamiltonian in the interaction representation
\begin{eqnarray}
H_{TI} &=&\frac{i\hbar\omega}{2} \Big[\Big(\beta
a\sigma_{+}e^{-i(\omega-E_{J}/\hbar)t}-\beta^{*}
a^{\dagger}\sigma_{-}e^{i(\omega-E_{J}/\hbar)t}\Big)\nonumber\\
&-&\Big(\beta^{*} a^{\dagger}\sigma_{+}e^{i(\omega+E_{J}/\hbar)t}-
\beta a\sigma_{-}e^{-i(\omega+E_{J}/\hbar)t}\Big)\Big]\nonumber\\
&-& E_z\Big(\sigma_{+}e^{E_{J}/\hbar t}+
\sigma_{-}e^{-E_{J}/\hbar t}\Big) \nonumber\\
&+& \frac{E_{J}}{2}\cos\Big[2\Big(\beta^{*}
 a^{\dagger}e^{i\omega t}+\beta ae^{-i\omega
t}\Big)+2\gamma \Big]\sigma_{z}\nonumber\\
&-& i\frac{E_{J}}{2}\sum_{n=0}^{\infty}\frac{{(2i)}^{2n}}{(2n)!}
\Big\{\Big\{\sin(2\gamma)e^{i\frac{E_{J}}{\hbar}t}+
\frac{2\cos(2\gamma)}{(2n+1)}\nonumber\\
&\times &\Big[\beta^{*} a^{\dagger}e^{i(\omega+E_{J}/\hbar)t}+
\beta ae^{-i(\omega+E_{J}/\hbar)t}\Big]\Big\}\nonumber\\
&\times &\Big[\beta^{* 2} a^{\dagger 2}e^{2i\omega t}+\beta^{2}
a^{2}e^{-2i\omega t}\nonumber\\
&+&|\beta|^{2}( a a^{\dagger}+ a^{\dagger} a) \Big]^{n}\sigma_{+} -
h.c\Big\}.\label{THTI}
\end{eqnarray}

\section{Resonance condition: Jaynes-Cummings type Hamiltonian}
At this stage we have in hands a transformed Hamiltonian equation
(\ref{THTI}) with a more complicated structure than the original
one, in equation (\ref{H}). Nevertheless, our new Hamiltonian may be
considerably simplified by choosing an appropriate resonance condition
and then applying the RWA. In fact, it is possible to obtain a well known
Hamiltonian of quantum optical resonance: the Jaynes-Cummings Hamiltonian
in the transformed frame.
If the cavity frequency is such that $\hbar\omega = E_{J}$ and  the
parameter $\gamma$ (which may be controlled by the classical flux)
equals $\gamma=\pi/4$ rad, the rapidly oscillating terms in the
right hand side of equation (\ref{THTI}) may be neglected (RWA), and
the transformed Hamiltonian above reduces to 
\begin{equation}
H_{TI,1} \approx -i\hbar(g^{*} a^{\dag}\sigma_{-}-g a\sigma_{+}),
\end{equation}
which coincides with the Jaynes-Cummings Hamiltonian with effective
coupling constant $|g| = |\beta|\omega/2$. Now, in order to allow
the RWA, the parameter $\beta$ cannot be arbitrarily large. In fact,
an analogue of the strong coupling regime requires that $|g| \ll \omega$,
or $|\beta| \ll 1$. Note that in the approach of reference \cite{yu04},
the condition $|\beta| \ll 1$ is also necessary, but for a different
reason, i.e., to truncate the co-sine (sine) series; see, for instance,
the discussion after equation (\ref{H}). We should remark
that in our scheme the Jaynes-Cummings evolution takes place in the
transformed frame, differently from the model developed in \cite{yu04}.

Now we would like to discuss some aspects of the dynamics of the 
system. Despite of the fact that
we have obtained a Jaynes-Cummings type Hamiltonian in the
transformed space, the system dynamics is closely related to the
dynamics of the driven Jaynes-Cummings model (DJCM), instead. In
this case, the time evolution of the state vector, for an initial
state $|\psi(0)\rangle$ is
\begin{equation}
|\psi(t) \rangle = T^\dagger {U}_{0T}(t){U}_I(t)T |\psi(0) \rangle,
\end{equation}
where ${U}_I(t)$ is the Jaynes-Cummings evolution operator
\cite{sten73} in the interaction representation

\begin{eqnarray}
{U}_I(t) & = & \frac{1}{2} \left[{C}_{n+1}+{C}_{n}\right]I +
\frac{1}{2} \left[{C}_{n+1}-{C}_{n}\right]
\sigma_{z}\nonumber\\
& + &\frac{\beta}{|\beta|} S_{n+1} a\sigma_{+}
-\frac{\beta^{*}}{|\beta|} a^\dagger S_{n+1}\sigma_{-},\label{U}
\end{eqnarray}
with ${C}_{n+1}=\cos\left(|g| t\sqrt{ a a^\dagger}\right)$ and $
S_{n+1}=\sin\left(|g| t\sqrt{ a a^\dagger}\right) /{\sqrt{ a
a^\dagger}}$.

Now we may calculate the qubit inversion
$W(t) = \langle\sigma_{z}\rangle$, having an initial state
$|\psi(0) \rangle=|e\rangle\otimes|\alpha_0\rangle$, i.e.,
the qubit in the excited state and the field in a coherent state
\cite{hofheinz08} with amplitude $\alpha_0$. The qubit inversion reads
\begin{eqnarray}
W(t)&=&\frac{1}{2}\exp(-|\alpha_0-\beta|^{2})
\sum_{n=0}^{\infty}\frac{|\alpha_0-\beta|^{2n}}{n!}\Big\{\nonumber\\
&&2\cos\Big(\frac{E_{J}}{\hbar}
t\Big)c_{n+1}c_{n}+\frac{n!}{\sqrt{n!(n+1)!}}\nonumber\\
&\times &\Big[(\alpha_0^{\star}-\beta^{\star})e^{i\frac{E_{J}}{\hbar}
t}+(\alpha_0-\beta)e^{-i\frac{E_{J}}{\hbar} t}\Big]\nonumber\\
&\times
&(c_{n+2}-c_{n})s_{n+1}-\frac{n!}{\sqrt{n!(n+2)!}}\Big[\nonumber\\
&(&\alpha_0^{\star}-\beta^{\star})^{2}e^{i\frac{E_{J}}{\hbar}
t}+(\alpha_0-\beta)^{2}e^{-i\frac{E_{J}}{\hbar}t}\Big]\nonumber\\
&\times &s_{n+2}s_{n+1}\Big\},\label{W}
\end{eqnarray}
where $c_{n}=\cos(gt\sqrt{n})$ and $s_{n}=\sin(gt\sqrt{n})$. In
equation (\ref{W}) the parameter $\beta$ was considered real for
simplicity.

\begin{figure}
\resizebox{0.95\columnwidth}{!}{%
  \includegraphics{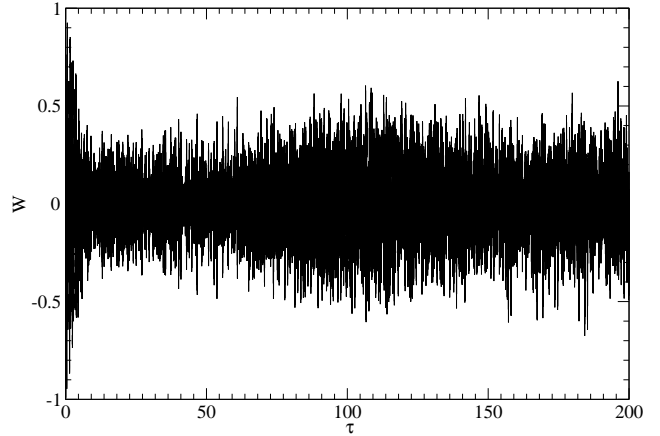}
}
\caption{Plot of the qubit inversion as a function of the
(dimensionless) scaled time $\tau=g t$. Ordinary revivals 
occurring for $\alpha_0=(5.0,0.5)$ and $\beta=0.02$.}
\label{fig1}       
\end{figure}

\begin{figure}
\resizebox{0.95\columnwidth}{!}{%
  \includegraphics{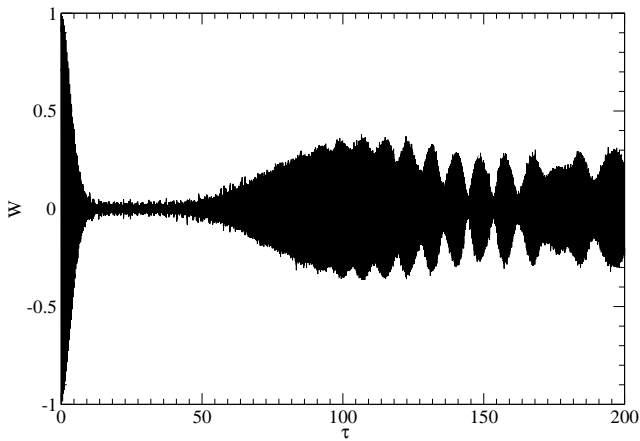}
}
\caption{Plot of the qubit inversion as a function of the
(dimensionless) scaled time $\tau=g t$. Super-revivals 
occurring for $\alpha_0=(0.5,5.0)$ and $\beta=0.02$.}
\label{fig2}       
\end{figure}

The structure of the equation above is similar to the one obtained
for $\langle\sigma_{z}\rangle$ in the DJCM \cite{hmcsmd94}.
Therefore we expect the phenomenon of super-revivals to be present
in the qubit-cavity system, analogously to the DJCM. Super-revivals
are revivals ocurring at larger time-scales than that of ordinary JCM
revivals, and which sometimes arise in the atom-field dynamics
\cite{hmcsmd94}.
This peculiar behavior is illustrated in figure (\ref{fig1}) and
figure (\ref{fig2}). We note that the existence or not of
super-revivals is narrowly connected to the preparation of the
initial field state. For instance, if we have $\alpha_0 =(5.0,0.5)$,
the super-revivals do not occur [see Fig. (\ref{fig1})], and we have
ordinary revivals only. However, for $\alpha_0 =(0.5,5.0)$, 
super-revivals take place in that system, as seen in Fig. (\ref{fig2}).

Regarding the influence of an external environment, we know that
in a non-ideal system, decoherence normally deteriorates quantum
effects. In this paper, though, we have studied the ideal system (without
an environment), in order to identify its main features. 
Nevertheless we would like to make a few comments about the non-ideal
situation. Real cavities have losses, which are usually modelled
by coupling the cavity field to an external thermal bath. A typical interaction term 
describing such a coupling may be written as (under the rotating wave 
approximation), $\sum_i \left(\lambda_i a^\dagger b_i + \lambda_i^* a b_i^\dagger\right)$, 
where the operators $b_i,b_i^\dagger$ refer to the environment. As our
transformation in Eq. (\ref{T}) contains displacement operators 
$D(\alpha,\gamma),D^\dagger(\alpha, \gamma)$ acting on the cavity field
sub-space, a term in the system-bath interaction having creation (annihilation)
field operators will transform basically as 
$D^\dagger(\alpha) a^\dagger D(\alpha) = a^\dagger + \alpha^*$
($D^\dagger(\alpha) a D(\alpha) = a + \alpha$ ). Thus, according to (\ref{T}),
the system-bath interaction part of the transformed Hamiltonian will contain 
terms of the type $a^\dagger b_i$, $\sigma_+ a^\dagger b_i$, and
$\sigma_z a^\dagger b_i$, for instance. Now, given that in our (interaction
representation) Hamiltonian only terms obeying a very
specific resonance condition will be relevant [see Eq. (\ref{THTI})], 
we do not expect extra contributions to the
effective master equation describing the non-ideal dynamics.
Generally speaking, for cavity having decay times of the order
of $10^{-3}$ s \cite{haroche01}, we believe that
super-revivals could be observed, as they would occur typically at a
much shorter time scale. In the example above, for instance, the
first super-revival occurs at $T_{sr}\approx 10^{-6}$ s, for a cavity
transition frequency $\omega = 10$GHz and $\beta = 0.02$.

\section{Conclusion}
In conclusion, we have presented a novel approach for studying the
dynamics of a superconducting qubit interacting with the quantized
field within a high-$Q$ cavity. In general, approximations
are made directly to the Hamiltonian. In
our method, we first apply an unitary transformation to the full
Hamiltonian in equation (\ref{H}) and make the relevant approximations 
after performing the transformation. Then, if a specific resonance condition
($\hbar\omega = E_{J}$) is chosen (as well as $\gamma=\pi/4$ rad) we
obtain a Jaynes-Cummings-type Hamiltonian after applying the RWA;
this constitutes the main result of our paper.
The comparison of our proposal with other approaches 
is not straightforward. Normally the
Hamiltonian is truncated after some kind of approximation - for instance,
by taking the limit $|\beta|<<1$, and more simple, linearized
\cite{yu04} or nonlinear Hamiltonians \cite{you03} are obtained. In
our method, we are able to obtain in a direct way, a Hamiltonian
which allows an exact solution for the state vector in a specific resonance
regime.
However, as the nonlinear effects are somehow
enclosed in the transformed Hamiltonian, they may give rise to a
more complex dynamics; in our example,
the resulting dynamics exhibits typical behavior of a driven
Jaynes-Cummings model \cite{hmcsmd94}
(or a trapped ion within a cavity \cite{moya}),
but without the presence of a classical driving field.
In particular, we have predicted the existence of
super-revivals for the superconducting qubit
inversion. We believe our approach could be useful
not only to establish a direct connection to other well known
models in quantum optics, but also the exploration of different
regimes in superconducting systems.
%
%
%
%
%
%

\vspace{0.5cm}

\noindent D.S.F thanks the financial support from Conselho
Nacional de Desenvolvimento Cient\'\i fico e Tecnol\'ogico-CNPq
(150232/2012-8), Brazil. We thank FAPESP and CNPq for financial
support through the National Institute for Science and Technology of
Quantum Information (INCT-IQ) and the Optics and Photonics Research
Center (CePOF).

\end{document}